\documentclass{jpsj2} 
\newcommand{\bi}[1]{\ensuremath{\boldsymbol{#1}}}
\title{Fano Effect in a Few-Electron Quantum Dot}

\author{Tomohiro \textsc{Otsuka}$^{1}$\thanks{t-otsuka@issp.u-tokyo.ac.jp},
Eisuke \textsc{Abe}$^{1}$,
Shingo \textsc{Katsumoto}$^{1}$,
Yasuhiro \textsc{Iye}$^{1}$, \\
Gyong L \textsc{Khym}$^{2}$, and
Kicheon \textsc{Kang}$^{2}$}

\inst{$^{1}$Institute for Solid State Physics,
University of Tokyo \\
5-1-5 Kashiwanoha, Kashiwa, Chiba 277-8581, Japan \\
$^{2}$Department of Physics and Institute for Condensed Matter Theory,\\
 Chonnam National University,
Gwangju 500-757, Korea}

\abst{We have studied the Fano effect in a few-electron quantum dot side-coupled to a quantum wire.
The conductance of the wire, which shows an ordinal staircase-like quantization without the dot,
is modified through the interference (the Fano effect) and the charging effects.
These effects are utilized to verify the exhaustion of electrons in the dot.
The ``addition energy spectrum" of the dot shows a shell structure,
indicating that the electron confinement potential is fairly circular.
A rapid sign inversion of the Fano parameter on the first conductance
plateau with the change of the wire gate voltage has been observed,
and explained by introducing a finite width of dot-wire coupling.}

\kword{quantum dot, quantum wire, Fano effect, charge detection, shell structure}

\begin{document}
\maketitle

\section{Introduction}
Electronic states of a quantum dot (QD) with only a few electrons resemble
those of a single atom, in that they both reveal, {\it e.g.},
shell structures and Hund's rule \cite{tarucha,tarucha3,tarucha2,2001KouwenhovenRPP}.
Such ``single-atom spectroscopy" of a QD is usually carried out by
measuring the transmission coefficient, that is, the conductance through it.
Such transmission measurement requires at least two connections to the wires.
There are two kinetic degrees of freedom in a quantum wire (QW):
the one along the wire with a continuum spectrum (longitudinal mode)
and the other crossing the wire with a discrete spectrum (transverse mode)\cite{Wees}.
In usual transmission experiments,
the longitudinal mode couples to the modes in the dot.
In order to realize a few-electron QD in so-called lateral type structures,
one needs to make the dot size small enough.
This size reduction usually suppresses the coupling either at the inlet or at the outlet of the QD,
and makes the transmission measurement in the few-electron regime difficult.
Though this difficulty is being overcome by recent advancement in gate design\cite{2000CiorgaPRB}
and optimized confinement voltages combined with built-in charge detectors\cite{2002SpinzakPRL,2003ElzermanPRB,2005KalishNat,2005JhonsonNat},
it still largely limits controllability of physical parameters.

In the side-coupled geometry illustrated in Fig.~\ref{fig_sem}(a) \cite{kob3,johnson,kang},
the QD should couple mainly with the transverse mode of the QW.
The negative voltage applied to the plunger gate (see Fig.\ref{fig_sem}(b)) ``pushes" the dot toward the wire,
and the QD is readily tuned to the few-electron regime,
still keeping a finite coupling to the wire.

It is noteworthy that the net current via the side-coupled QD is zero.
Nevertheless, the effect of the QD appears in the conductance of the QW indirectly through two effects:
the interference effect and the potential modulation effect.
The former arises from the interference between the path
directly passing through the QW and the path having a bounce to the QD \cite{kang,johnson,kob3}.
Around a resonant energy $\epsilon_0$, 
the complex transmission probability $t(\epsilon)$ through the QD 
for a conduction channel with energy $\epsilon$
exhibits the Breit-Winger form \cite{breit}:
\begin{equation}
t(\epsilon)\propto\frac{1}{(\epsilon-\epsilon_0)+i\gamma},
\label{breit-wigner}
\end{equation}
where $\gamma$ is the width of the resonance.
It is apparent from Eq. (\ref{breit-wigner}) that
the phase shift through the QD varies by $\pi$ when $\epsilon$ crosses $\epsilon_0$.
But in the reflection path, the electron goes through the QD twice during a single scattering event,
which results in the phase shift $\Delta\theta=2{\rm arg}(t(\epsilon))$.
Such resonant transmission with the interference is
generally described by the Fano formula \cite{fano,entin,kob1,kob2}:
\begin{equation}
F(E)\propto\frac{(E-q)^2}{{E}^2+1},
\label{fano_lineshape}
\end{equation}
where the transmission probability $F$ is expressed as a function of
normalized energy $E=(\epsilon-\epsilon_0)/\gamma$ and the Fano parameter $q$.
In the case of the side-coupled QD with an ideal single-point contact,
$q$ should be zero as long as we ignore the two-dimensionality of the dot.
In reality, the contact may have finite widths and the QD may have two-dimensional structures,
which allows $q$ to be non-zero.

In the side-coupled structure, the QD and the QW are so close that 
the electrostatic potential of the QD affects the QW. 
The change of the electron number in the dot from $N$ to $N \pm 1$
results in a rapid change of the potential,
thereby the conductance through the QW also changes abruptly\cite{1993FieldPRL,2002SpinzakPRL,2003ElzermanPRB,2005KalishNat,2005JhonsonNat}.
This effect is particularly enhanced
when the gate voltage of the QW is tuned around a transition region,
where the conductance is most sensitive to the surrounding potential.

In this article, we present an experimental study on the transport
through a QW with a side-coupled QD in the few-electron regime.
We demonstrate, from the observed charging and interference effects,
that $N$ is indeed reduced to zero.
The ``addition energy spectrum" shows the shell structure
due to two-dimensional harmonic potential,
indicating that the confinement potential possesses highly circular symmetry.
On the first plateau of the conductance quantization of the QW,
we find that the Fano parameter rapidly changes its sign with the change of the wire gate voltage.
We present simple models to explain this observation.
The models take a finite width of the dot-wire contact into account, and reasonably reproduce our experimental results.

\section{Experiment}
\begin{figure}
\begin{center}
(a)\includegraphics[width=0.2\linewidth]{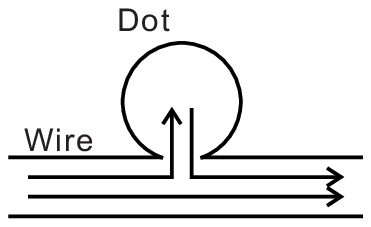}
(b)\hspace{2mm}\includegraphics[width=0.2\linewidth]{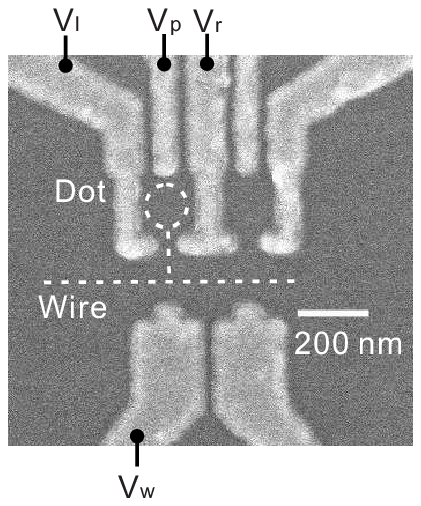}
\end{center}
\caption{(a) Schematic view of a side-coupled geometry.
(b) Scanning electron micrograph of a sample similar to the one measured.
White regions are metallic gates.
The broken circle and line indicate the positions of the QD and the QW, respectively.
}
\label{fig_sem}
\end{figure}

The side-coupled system used in the present experiment was made
from an AlGaAs/GaAs heterointerface buried 60 nm below the surface.
The dot and the wire were defined by Ti/Au metallic gates.
The gate configuration is shown in Fig.~\ref{fig_sem}(b).
Though this sample was designed to have two QDs in parallel,
only the left dot in the figure was used in the present study.
The electrostatic potential of the dot was controlled by 
the plunger gate voltage $V_{\rm p}$, and the wire width was controlled
by the wire gate voltage $V_{\rm w}$.

The sample was cooled to 30 mK by a dilution refrigerator.
The wire conductance was measured by a standard lock-in technique
with the frequency of 80 Hz.

\section{Results and discussion}
\subsection{Addition energy spectrum}
As is well-known, the conductance of a perfect QW
is quantized with the conductance quantum $e^2/h$.
The conductance is simply written as $2e^2M/h$,
where $M$ is the number of conducting channels and
the factor 2 comes from the spin degeneracy.
The wire gate voltage $V_{\rm w}$ controls $M$.
Consequently,  a staircase-like conductance lineshape is seen as a function of  $V_{\rm w}$\cite{Wees}.
The effects of the QD are superposed on this staircase conductance as perturbations.

\begin{figure}
\begin{center}
\includegraphics{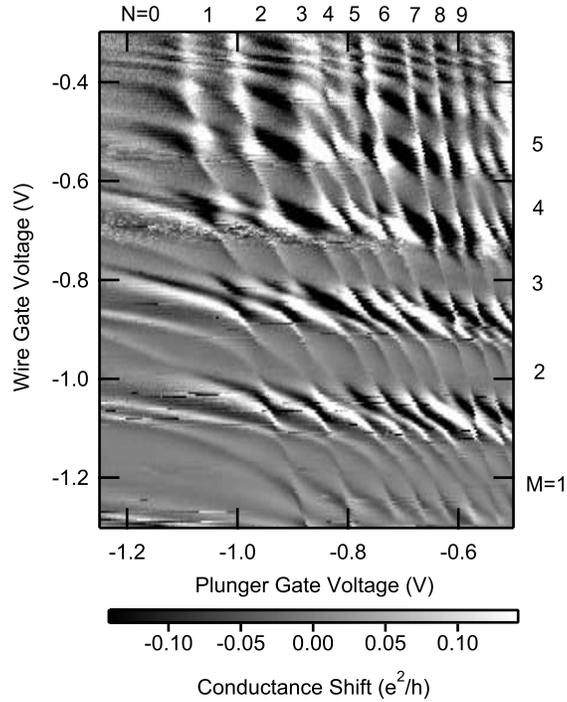}
\end{center}
\caption{Gray scale plot of the wire conductance shift from ordinal stepwise conductance of the QW,
as a function of the dot gate voltage and the wire gate voltage.
$M$ is the index of wire conductance quantization and
$N$ is the number of electrons in the dot.
Fano peaks appear as strong black-white stripes.
}
\label{fig_otsuka}
\end{figure}

In order to extract the effects of the QD, we adopt the following procedure.
The perturbations by the QD should be driven by $V_{\rm p}$,
while it should affect the wire conductance only slightly
thanks to the longer distance between the dot plunger gate and the wire (Fig.\ref{fig_sem}(b)).
Hence the conductance lineshape as a function of the plunger gate voltage $G(V_{\rm p})$
at a fixed value of $V_{\rm w}$ contains the perturbations by the QD as rapidly varying components.
On the other hand, the primary component should be immune to $V_{\rm p}$ and 
the direct electric field effect should cause a slow modulation on it.
We subtracted the latter two by fitting a slowly varying function
to $G(V_{\rm p})$ and obtained the residual perturbation term $\Delta G(V_{\rm p})$.
Repeating this procedure for different values of $V_{\rm w}$, 
we obtained the shift of the conductance $\Delta G$ due to the interference and charging perturbations.
$\Delta G$ as a function of $V_{\rm p}$ and $V_{\rm w}$ is shown in Fig.~\ref{fig_otsuka}.

\begin{figure}
\begin{center}
\includegraphics{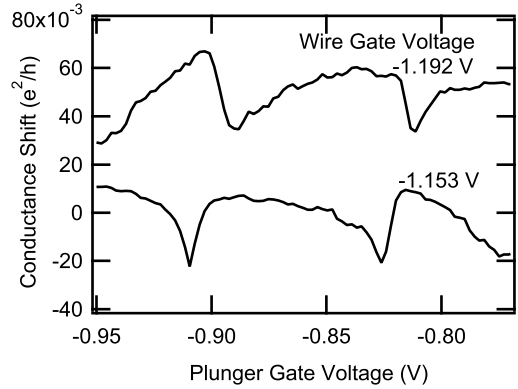}
\end{center}
\caption{Conductance shift $\Delta G$ is plotted as a function of  $V_{\rm p}$
at $V_{\rm w}$=$-$1.153 V and $-$1.192 V.
The data for $V_{\rm w}$=$-$1.192 V are shifted by 0.05$e^2/h$ for clarity. 
These two wire gate voltages are in the region of $M$=1.
Note that the direction of the distortion is reversed
in spite of the small change of $V_{\rm w}$.
}
\label{fig_fano_lineshape}
\end{figure}

The amplitude of $\Delta G$ is apparently larger
at the boundaries between different $M$,
due to the high sensitivity of charge fluctuation in the transition regions.
Black and white stripes flow vertically with slanting.
These duotone lines correspond to the Fano lineshape given in Eq.(\ref{fano_lineshape}),
each of which has a peculiar peak and dip structure.
Since they arise from resonance in the QD,
a cross section of Fig.~\ref{fig_otsuka} along a constant $V_{\rm p}$ contains
information on the energy spectrum
analogous to the Coulomb oscillation in a conventional transmission experiment.
Examples for $V_{\rm p}$= $-$1.153 V and $-$1.192 V 
are shown in Fig.~\ref{fig_fano_lineshape}.

The assignment of the number of electrons in the dot $N$ is 
the first step of the analysis in the few-electron regime.
For this purpose, it is usually requested to find the Coulomb valley at $N=0$.
In the present case, the assignment is clearly made as follows;
The duotone lines corresponding to the Fano resonance disappear in the region $V_{\rm p}<-$1 V,
which we assign to the region of $N=0$.
(We note that the remaining horizontal lines in this region are not Fano resonances. 
We will come back to this issue in the last part of this paper.)
In a conventional two-contact transmission experiment, a steep decrease 
of the Coulomb peak heights with decreasing $N$ obstruct
the distinction between the exhaustion of electrons and the detachment of the dot from the electrodes.
On the contrary, the side-coupled structure is free from this isolation problem,
since the wire and the dot have only one connection as we explained above.
This is supported by the observation that the black-white contrast of the duotone lines
does not decrease with decreasing $N$,
and that no new duotone line emerges in the $N=0$ region,
even the coupling between the QD and the QW is strengthened.
Furthermore, the strong oscillations at the boundaries between adjacent $M$'s
rapidly drop to small values for $V_{\rm p}<-1$ V.
This is equivalent to a charge sensing experiment and certifies
that the Coulomb peak does not exist in this region.
The assignment of $N$ is displayed at the top of Fig.~\ref{fig_otsuka}.

\begin{figure}
\begin{center}
\includegraphics{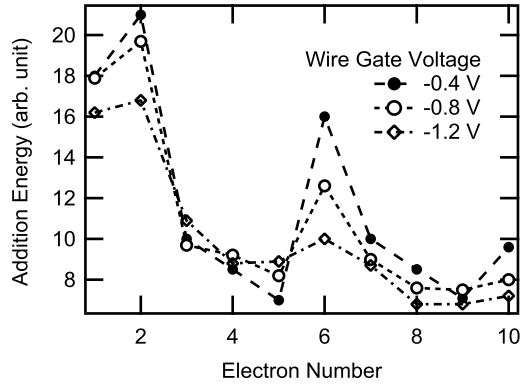}
\end{center}
\caption{
Addition energy spectrum of the dot obtained from the distances
between the Fano peaks for three different wire gate voltages.}
\label{fig_addition}
\end{figure}

The energy spectrum of the QD appears in the distances between the peaks.
The energy required to add one electron to the QD (``addition energy")
is mainly the sum of the single-electron charging energy and the kinetic (orbital) energy.
Here, we ignore the correlation or other many-body effects and apply
a simplest ``constant capacitance" approximation to the interaction between electrons,
in which the addition of one electron is described
by the constant energy gain of  $E_{\rm c}$
(electrostatic charging energy )\cite{2001KouwenhovenRPP}.
We also assume that the gate voltage $V_{\rm p}$ just linearly shifts
the electrostatic potential of the dot.
Under a resonant condition, the electrochemical potentials in the dot and wire are balanced.
Then the energy shift required to reach the next resonant level
is the sum of the energy difference in the single-electron orbital energy
$\Delta\epsilon_{\rm orb}$ and the charging energy $E_{\rm c}$.
$\Delta\epsilon_{\rm orb}$ is zero when the next electron occupies
an orbital state degenerate to the previous one.
Hence, in this approximation, the smallest distance corresponds to
the pure electrostatic charging energy and the enhancement from the value gives $\Delta\epsilon_{\rm orb}$.
Figure~\ref{fig_addition} shows the addition energy (the distances of the Fano lines) as a function
of $N$ for various $V_{\rm w}$ in arbitrary unit.
Unfortunately, the direct current-voltage characteristics of the dot 
cannot be carried out in the present structure.
This complicates the estimation of the addition energy with high reliability.
Therefore, we only discuss the relative amplitudes of the addition energy here.
In Fig.~\ref{fig_addition},
apparent peaks appear at $N=2$ and $6$, 
at which the system is more stable than the neighboring ones.
These numbers correspond to the first two
shells of the two-dimensional symmetric harmonic oscillator ({\it i.e.}, $n(n+1)$ for $n=1,2$).
For the stable shell at $N=6$ to be formed,
the potential profile should be harmonic, and that the potential should be isotropic (circular).
When $N$ exceeds $10$, the constant capacitance model apparently breaks down,
and monotonic decrease in the peak distance is observed.

At present we do not have a clear interpretation to the relative heights of 
the peaks in the addition energy, which are more prominent for lower $|V_{\rm w}|$,
{\it i.e.}, higher wire conductance.
One possible interpretation is that the narrowing of the wire distorts
the circular shape of the potential into an elliptic one.
The effect of electrostatic potential of $V_{\rm w}$ on 
the dot potential is observed in the oblique angle of the 
Fano duotone lines in Fig.~\ref{fig_otsuka}.
However, this process only breaks the degeneracy of orbital levels
by unequal enhancement\cite{1994MadhavPRB,1999AustingPRB},
hence this effect does not solely explain the lowering of the peak at $N=2$,
which originates from the Kramers degeneracy.
Another possibility is the effect of the transverse modes in the wire.
In a real system, only the transverse modes can penetrate into the dot and
the coupling between the transverse modes and the orbitals in the dot
should cause the lifting of the level degeneracy.
At a plateau of conductance $2Me^2/h$, the $M$th mode mainly couples
to the dot because of the highest transverse energy.
This level mixing effect may have nontrivial dependence on
$V_{\rm w}$ due to the difference in the spatial configurations of the
confinement potentials and may result in the observed dependence of
the prominence on $V_{\rm w}$.

\subsection{Fano lineshape on the first plateau}
In this subsection, we focus on the Fano lineshapes on the first conductance plateau ($M=1$),
where the interference effect dominates the lineshape.
Because we are observing the Fano effect in the longitudinal transport,
there should occur transverse-longitudinal mixing at the dot-wire contact point.
The mixing at a single point is naturally described by a single S-matrix.
Then the Fano parameter $q$ is zero.
The fact is that, as clearly demonstrated in Fig.~\ref{fig_fano_lineshape},
the lineshapes reveal non-zero $q$, indicating some ``spatial structures" at the connecting point.

\begin{figure}
\begin{center}
\includegraphics{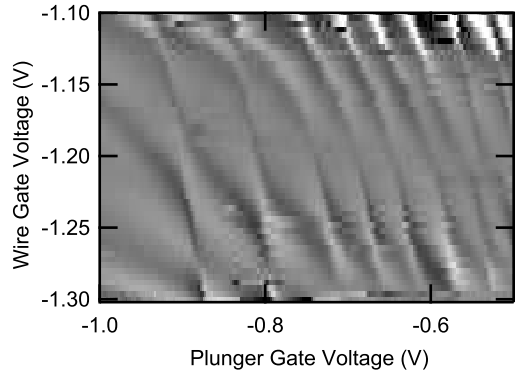}
\end{center}
\caption{Close up of Fig.~\ref{fig_otsuka}
around the first ($M$=1) conductance plateau.
Gray scale is the same as that in Fig.~\ref{fig_otsuka}.
}
\label{fig_q_change}
\end{figure}

In Fig.~\ref{fig_fano_lineshape},
the sign of $q$ changes upon a small change of $V_{\rm w}$.
This is also seen from Fig.~\ref{fig_q_change},
in which the duotone lines are twisted around the midway of the $M=1$ plateau.
Because $V_{\rm w}$ varies the longitudinal wave vector on the conductance plateaus,
this observation manifests that
the dot-wire connecting point must have a finite width as we explain below.

\subsection{A model for the coupling between the QW and the QD}
\begin{figure}
\begin{center}
\includegraphics[width=0.35\linewidth]{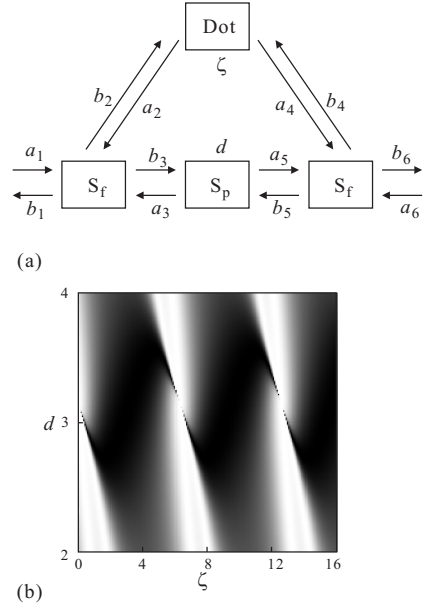}
\end{center}
\caption{(a) Theoretical model of a side-coupled dot system. A single channel is assumed in the wire.
Rectangles represent S-matrices. ${\rm S}_{\rm f}$'s are fork matrices, which are identical to $\bi{S}_{\rm T}$ in
Eq.(\ref{SmatrixT}).
The dot S-matrix is composed of two barrier matrices ($\bi{S}_{\rm D}$ in Eq.(\ref{barrier_smatrix}))
and a phase shifter with the phase shift $\zeta$
(Note that $\zeta$ is the inner phase shift and the total shift of the dot is different.)
${\rm S}_{\rm p}$ is also a phase shifter with a phase shift of $d$.
(b) Gray-scale plot of calculated transmission of the system in (a) as a function of phase shifts $\zeta$ and $d$.
White corresponds to 1 and black to 0 transmission probability. Note the similarity to Fig.~\ref{fig_q_change}.}
\label{fig_mpoint_sim}
\end{figure}

The above-mentioned behavior is reproduced by simple theoretical models,
in which a finite width of the contact along the QW is taken into account.
The simplest among them may be the triangular circuit model illustrated in
Fig.~\ref{fig_mpoint_sim}(a).
Within the framework of scattering formalism,
three-leg S-matrix may be chosen as\cite{datta}
\begin{gather}
\left(
\begin{array}{l}
b_1\\ b_2\\ b_3
\end{array}
\right)
=\bi{S}_{\rm T}
\left(
\begin{array}{l}
a_1\\ a_2\\ a_3
\end{array}
\right)
,\\
\;\;\;\;\;
\bi{S}_{\rm T}=
\left(
\begin{array}{ccc}
\frac{1-a}{2} & -\frac{1+a}{2} & \sqrt{\frac{1-a^2}{2}}\\
-\frac{1+a}{2} & \frac{1-a}{2} & \sqrt{\frac{1-a^2}{2}}\\
\sqrt{\frac{1-a^2}{2}} & \sqrt{\frac{1-a^2}{2}} & a
\end{array}
\right)
,
\label{SmatrixT}
\end{gather}
for the two vertices to keep the unitarity.
Here, we take $a$ to be a real number, which determines the direct reflection coefficient.
The S-matrix for a two-leg vertex is generally expressed as
\begin{gather}
\left(
\begin{array}{l}a_3'\\ a_4\end{array}\right)
=\boldsymbol{S}_{\rm D}
\left(\begin{array}{l}b_3' \\ b_4\end{array}\right),
\;\;\;\;\;\\
\boldsymbol{S}_{\rm D}=
\left(
\begin{array}{cc}
\cos\phi & e^{i\beta}\sin\phi\\ e^{i\beta}\sin\phi & \cos\phi
\end{array}\right),
\label{barrier_smatrix}
\end{gather}
where $\phi$ determines the reflection coefficient and
$\beta$ corresponds to the phase shift.
When the reflection coefficient is large,
this matrix works as a ``barrier" for electrons.
For $\beta=\pi/2$ on the other hand, there is no reflection and
the matrix works simply as a ``phase shifter".
In an electron waveguide, $\beta$ is the product of
the wave vector $k$ and the length $L$.
Hence a QD can be expressed as a phase shifter (waveguide) sandwiched by
two barriers if the charging energy is ignored.
The gate voltage $V_{\rm p}$ varies $k$ in the dot and causes resonances.
In the model illustrated in Fig.~\ref{fig_mpoint_sim}(a),
the QD is approximated as above and
the QW can be expressed as a phase shifter,
in which the phase shift $d$ monotonically depends on the gate $V_{\rm w}$.

Figure~\ref{fig_mpoint_sim}(b) shows calculated conductance based on this model
as a function of $d$ (the phase shift  in the wire ) and $\zeta$ (the phase shift {\it inside} the dot).
The phase shifts $d$ and $\zeta$ should be
monotonic functions of $V_{\rm w}$ and $V_{\rm p}$, respectively.
Therefore, as a crude approximation,
we may directly compare Fig.~\ref{fig_q_change} and Fig.~\ref{fig_mpoint_sim}.
Despite the simplicity of the present model,
the calculation shows similarity to the experiment.
This demonstrates that the model captures essential ingredients in the system.
Note that the phase shift in the wire depends on the wave vector $k$
and the above simple picture of single conduction channel is applicable only to the $M=1$ plateau.
However, the Fano lineshapes arise from the resonance at the Fermi level,
and in general the resonant condition selects a single $k$\cite{kob3}.
This is the reason that similar phenomena are also visible on the higher conductance plateaus.
Unfortunately, they are not so clear as that on the $M=1$ plateau,
presumably due to the multi-band effect.

\begin{figure}
\begin{center}
(a)\includegraphics[width=0.35\linewidth]{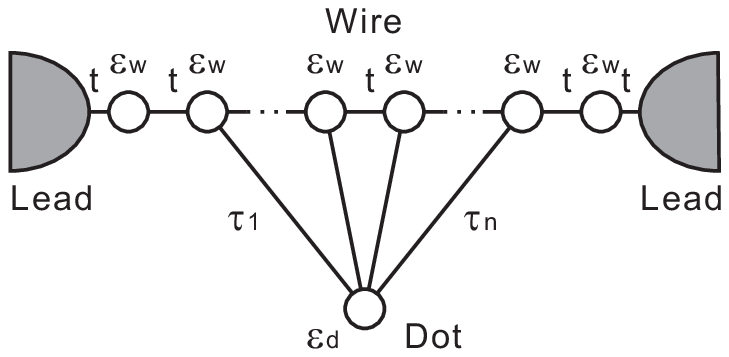}\\
(b)\includegraphics{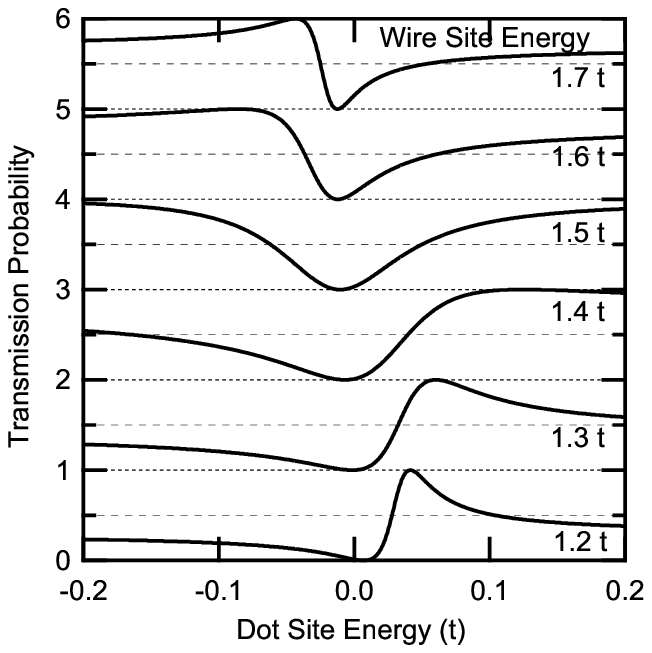}\\
\end{center}
\caption{(a) Theoretical model for a finite width contact
based on the tight-binding approximation.
(b) Fano lineshape and its evolution with the change of the wire gate voltage. This result is
calculated using the model illustrated in (a).
Here we take $n=8$ and $\tau_1=\tau_8=0.1t$, $\tau_2=\cdots=\tau_7=0$.
}
\label{fig_kang_theory}
\end{figure}

Another model can be considered within the tight-binding approximation.
An example is illustrated in Fig.~\ref{fig_kang_theory}, in which the total
Hamiltonian $H$ is given as $H=H_\nabla+H_L+H_R+H_T$,
where $H_\alpha$ is the Hamiltonian for leads ($\alpha=L,R$ referring to the left and right leads),
$H_T$ for the coupling to the leads, and $H_\nabla$ for the $\nabla$-shaped central part.
The QW part is expressed as a finite chain of $n$ sites.
Each site has a same energy $\epsilon_{\rm w}$,
and the neighboring sites are connected with a coupling constant $t$.
The QD is expressed as a localized site with an energy $\epsilon_d$.
The QD site couples to the $i$th site with an energy $\tau_i$.
Hence $H_\nabla$ is written as
\begin{equation}
\begin{aligned}
H_\nabla=\sum_{i=1}^n\epsilon_ic_i^\dagger c_i
&-\sum_{i=1}^{n-1}t(c_{i+1}^\dagger c_i+c_i^\dagger c_{i+1})\\
&-\sum_{i=1}^n\tau_i(c_d^\dagger c_i+c_i^\dagger c_d)
+\epsilon_dc_d^\dagger c_d.
\end{aligned}
\end{equation}

Calculation of the transmission probability based on this model is straightforward.\cite{datta}
The results are shown in Fig.~\ref{fig_kang_theory}(b)
as a function of $\epsilon_d$ for several $\epsilon_w$ (in unit of $t$).
As expected, the results reproduce rapid change in $q$ with a small change in $\epsilon_w$.
In this model, the sign-inversion of $q$ occurs at the band of the wire.
In other words, the dips and peaks trace ``anticrossing-like lines" along the states in the dot.
This gives another method to detect the energy level evolution with gate voltages.
For example, in the region assigned as $N=0$ in Fig.~\ref{fig_otsuka}, 
such black ({\it i.e.}, dip) traces move slowly upward with lowering $V_{\rm p}$,
while they merge into the Fano lines with nonzero $N$.
This observation again certifies the assignment of $N$.

In summary, we have studied the Fano effect in a quantum dot side-coupled to a quantum wire.
In particular, the dot was tuned to the few-electron regime, and
the $N=0$ state has been identified both from the charge detection
and disappearance of the Fano lineshape.
The addition energy spectrum shows a clear shell structure,
indicating that the electron confinement potential is circular and symmetric.
A rapid sign inversion of the Fano parameter on the first conductance
plateau has been observed, and explained by introducing a finite width
of the dot-wire contact.

\section*{Acknowledgments}
The authors thank M. Eto, who suggested the possibility of 
observation of the interference effect down to zero-electron
in the side-coupled configuration.
This work is supported in part by a Grant-in-Aid for Scientific Research from the Japanese
Ministry of Education, Culture, Sports, Science and Technology and also partly supported by Special Coordination
Funds for Promoting Science and Technology.

\end{document}